\documentclass[aps,prc,showpacs]{revtex4-1}
\usepackage[dvips]{graphicx}
\usepackage{amsmath}
\usepackage{cases}
\begin{document}
%%%%%\draft
\title{Description of a domain by a squeezed state in a scalar field theory}
\author{Masamichi Ishihara}
\thanks{Electric address: m\_isihar@koriyama-kgc.ac.jp}
\date{\today} 
%%%%%% abstract
\begin{abstract}
The author attempted to describe a domain by using a squeezed state in quantum field theory. 
An extended squeeze operator was used to construct the state. 
In a scalar field theory, the author described a domain 
that the distributions of the condensate and of the fluctuation are Gaussian.
The momentum distribution, chaoticity and correlation length were calculated.
It was found that 
the typical value of the momentum is about the inverse of the domain size, 
and that 
the chaoticity reflects the ratio of the size of the squeeze region to that of the coherent region.  
%%%
The results indicate that 
the quantum state of a domain is surmised by these quantities under the assumption 
that the distributions are Gaussian.
As an example, this method was applied to a pion field,
and the momentum distribution and the chaoticity were shown.
%%%%%%%%%%%%%%%%%%%%%%%%%%%%%%%%%%%%%%%%%%%%%%%%%%%%%%%%%%%%%%%%%%%%%%%%%%%%%%%%%%%%%%%%
\end{abstract}

\maketitle
%%%%%%%%% Introduction %%%%%%%%%
\section{Introduction}
\label{sec:intro}
Squeezing is a basic concept for various branches of physics 
and is applicable for many subjects. 
This concept is used in quantum optics \cite{book:squ,Yuen,Walls}, 
quantum field theory, etc.
The squeezed state is a good base to study the non-perturbative nature 
in quantum mechanics and quantum field theory.
For example, this state appears in particle creation 
\cite{production_Parker,production_Boyanovsky} and 
is used to calculate effective potential 
\cite{EPSQS_H_Mishra, EPSQS_A_Mishra_56, EPSQS_A_Mishra_23, EPSQS_Tdel}.
The important property is the relation of fluctuations of conjugate variables. 
Only a coherent state does not minimize the uncertainty relation, 
but a squeezed state also does.

%%%%% 
Domains will be formed in some processes.
For instance, it is expected that a domain is formed in a chiral phase transition. 
The condensate is a spatial dependent in such a phenomenon,
and the spatial dependence indicates a domain formation.
The condensate is described by a coherent state which is constructed by a displacement operator\cite{our}.

Fluctuation is also an important quantity and is related to mass modification.
It was pointed out that mass modification is related to particle correlation
\cite{Asakawa_PRL, Sandra_BBC_2006,Csorgo_BBC_2007}. 
The fluctuation will be spatial dependent on a domain like condensate in some systems,
and a squeezed state is a candidate to describe the fluctuation. 
The naive extension of one mode squeeze operator in quantum mechanics
does not work well in quantum field theory, because of infinite volume.
The divergence is easily shown when the state constructed by the squeeze operator is used. 
This divergence is removed 
by introducing smeared creation and annihilation operators \cite{contsqu}. 
%%%%%%%%%%%%%%%%
It is worthwhile to find the squeezed state that 
describes a spatial dependent fluctuation without the divergence caused by infinite volume,  
because this state is related directly to observables.
Physical quantities will be calculated when the squeezed state is given. 

%%%%%%%%%%%%%%%%
In this paper, we attempt to describe a domain by a squeezed state without the divergence.
We use an extended squeeze operator,
and construct the squeezed state to describe a spherical domain:
the distributions of the condensate and of the fluctuation are Gaussian.
The momentum distribution, chaoticity and correlation length are calculated by using the obtained squeezed state.
It is demonstrated that 
the momentum distribution reflects the size of the domain, 
and that 
the chaoticity reflects the ratio of the size of the squeezed region 
to that of the coherent region.
Contrarily, 
the results imply that 
the quantum state describing a domain is surmised by studying the momentum distribution 
and the chaoticity
under the assumption that the distributions of the condensate and the fluctuation are Gaussian.

%%%%%%%%%%
This paper is organized as follows. 
In Sec.~\ref{sec:sqop}, 
an extended squeeze operator is introduced and a basic relation (Bogoliubov transformation) is derived. 
In Sec.~\ref{sec:domain}, 
we describe a spherical domain by a squeezed state.
The spatial configuration of the condensate and that of the fluctuation are taken 
to be Gaussian.
The momentum distribution, chaoticity, and correlation length are calculated.
Section~\ref{sec:conclusion} is assigned for conclusion and discussion.

%%%%%%%%%%%%%%%%%%%%%%%%%%%%%%%%%%%%%
\section{Squeeze Operator}
\label{sec:sqop}
A coherent state is a useful tool to represent electro-magnetic field in quantum optics.
The state is an eigenstate of an annihilation operator and is a minimum uncertain state. 
A squeezed state is also an minimum uncertainty state and 
is characterized by its fluctuation and shows non-classical behaviors. 
This state is applied to many topics in quantum field theory
like particle production in early universe. 
Only two modes, $\vec{p}$ and $-\vec{p}$, couple each other,
when the squeezing occurs in an infinite system.
Then two mode squeeze operator is the enough tool to represent the state. 
However, it is difficult to represent the state of squeezing for a finite object 
by two mode squeeze operator, because many mode couples.

The displacement operator is used to describe a coherent state and is defined as follows:
\begin{equation}
D(\alpha) = \exp \left ( \int d\vec{k} 
\left[ \alpha(k) a^{\dag}(k)  - \alpha^{*}(k) a(k) \right] \right),
\label{def:disp}
\end{equation}
where $\vec{k}$ represents momentum, $a(\vec{k})$ is an annihilation operator 
which satisfies the commutation relation: 
$[a(\vec{k}),a^{\dag}(\vec{l})] = \delta(\vec{k}-\vec{l})$ and 
$[a(\vec{k}),a(\vec{l})] = 0$.

It seems that a squeeze operator is naively extended by referring to one mode squeeze operator: 
\begin{equation}
\tilde{S}(\epsilon) = 
\exp \left(\frac{1}{2} \int d\vec{k} \left\{ 
\epsilon^{*}(\vec{k}) \left[a(\vec{k}) \right]^{2} 
- \epsilon(\vec{k}) \left[ a^{\dag}(\vec{k}) \right]^{2} 
\right\} \right) .
\label{eqn:unsatisfact_op}
\end{equation}
However this does not work well
because the squeezed state constructed with this operator generates divergence \cite{contsqu}:
%%%%%%%%%%%%%%%%%%%%%%%%%%%%%%%%%%%%%%%%%%
\begin{eqnarray}
\langle 0 | \tilde{S}^{\dag}(\epsilon) 
\left( \int d\vec{k} \ \omega(\vec{k}) a^{\dag}(\vec{k})  a(\vec{k}) \right)
\tilde{S}(\epsilon) | 0 \rangle
= \delta(0) \int d\vec{k} \ \omega(\vec{k}) \sinh^{2} \left( \left| \epsilon(\vec{k}) \right| \right)
, 
\label{eqn:div}
\end{eqnarray}
where $\omega(\vec{k})$ is the energy of a free particle
and the state $|0\rangle$ is the vacuum that is defined by $a(\vec{k}) |0\rangle =0$.
Eq.~\eqref{eqn:div} is apparently diverge except for
$\sinh  \left( \left| \epsilon(\vec{k}) \right| \right) \equiv 0 $.

Therefore we adopt the following squeeze operator by referring to two mode squeeze operator:
\begin{equation}
S(G) = \exp \left( \frac{1}{2} \int d\vec{k} d\vec{l}
\left[ G^{*}(\vec{k},\vec{l}) a(\vec{k}) a(\vec{l}) 
- G(\vec{k},\vec{l}) a^{\dag}(\vec{k}) a^{\dag}(\vec{l}) \right] 
\right), 
\label{def:contS}
\end{equation}
where $a(\vec{k})$ is the annihilation operator
and $G(\vec{k},\vec{l})$ is symmetric: $G(\vec{k},\vec{l}) = G(\vec{l},\vec{k})$. 
The displacement operator, eq.(\ref{def:disp}), 
and the squeeze operator, eq.(\ref{def:contS}), 
have following relations:
%%%%%%%
\begin{subequations}
\begin{align}
D^{\dag}(\alpha)a(\vec{p})D(\alpha) &=  a(\vec{p}) + \alpha(\vec{p}), \\ 
%%%%%%
S^{\dag}(G) a(\vec{p})S(G) &= a(\vec{p}) 
%%%
+\sum_{n=1}^{\infty} \frac{1}{(2n)!}
\int d\vec{k}_{2n} A_{2n}(\vec{p},\vec{k}_{2n}) a(\vec{k}_{2n})
\nonumber \\ & \qquad 
- \sum_{n=0}^{\infty} \frac{1}{(2n+1)!}
\int d\vec{k}_{2n+1} B_{2n+1}(\vec{p},\vec{k}_{2n+1})
a^{\dag}(\vec{k}_{2n+1})
%%%
, \label{eqn:sop}
\end{align}
\end{subequations}
%%%%%%%
where $A_{2n}$ and $B_{2n+1}$ are  
%%%%%%%
\begin{subequations}
\begin{align}
&
A_{2n}(\vec{p},\vec{k}_{2n}) := 
\int d\vec{k}_{1} \cdots d\vec{k}_{2n-1} 
G(\vec{p},\vec{k}_{1}) G^{*}(\vec{k}_{1},\vec{k}_{2}) 
G(\vec{k}_{2},\vec{k}_{3}) \cdots G^{*}(\vec{k}_{2n-1},\vec{k}_{2n}), \\
&
B_{2n+1}(\vec{p},\vec{k}_{2n+1}) :=
\int d\vec{k}_{1} \cdots d\vec{k}_{2n} 
G(\vec{p},\vec{k}_{1}) G^{*}(\vec{k}_{1},\vec{k}_{2})
\cdots G(\vec{k}_{2n},\vec{k}_{2n+1}).
\end{align}
\end{subequations}
%%%%%%%%%%%%%%%%%%%%%%%%%%%%%%%%%%
The squeezed state $|{\rm sq} \rangle$ is defined with $D(\alpha)$ and $S(G)$:
\begin{equation}
\left| {\rm sq} \right\rangle = D(\alpha) S(G) \left| 0 \right\rangle .
\label{eqn:def:sqs}
\end{equation}
%%%%%%%%%%%%%%%%%%%%%%%%%%%%%%%%%%

It is difficult to deal with this squeezed state, eq.~\eqref{eqn:def:sqs}, without any assumption.  
Therefore we use a decoupling type of $G(\vec{k},\vec{l})$ in this study: 
\begin{equation}
G(\vec{k},\vec{l}) = g(\vec{k})  g(\vec{l}) .
\end{equation}
The functions $A_{2n}\left( \vec{p},\vec{k}_{2n} \right) $ and 
$B_{2n+1}\left( \vec{p},\vec{k}_{2n+1} \right)$ are given as follows:
\begin{subequations}
%%%%%%%
\begin{align}
& A_{2n}\left( \vec{p},\vec{k}_{2n} \right) = L^{2n-1}  g(\vec{p})  g^{*}(\vec{k}_{2n}) ,
\\ 
& B_{2n+1}\left( \vec{p},\vec{k}_{2n+1} \right) = L^{2n}  g(\vec{p})  g(\vec{k}_{2n+1}),
\end{align}
\end{subequations}
where ${\displaystyle L=\int d\vec{k} \ g(\vec{k}) g^{*}(\vec{k})}$. 
%%%%%%
The transformation of the annihilation operator by the squeeze operator is 
\begin{equation}
S^{\dag}(G) a(\vec{p}) S(G)  = 
a(\vec{p}) + 
\left( \frac{\cosh(L) - 1 }{L}\right) 
g(\vec{p}) \left( \int d\vec{k} g^{*}(\vec{k}) a(\vec{k}) \right) 
- \left( \frac{\sinh(L)}{L} \right) 
g(\vec{p}) \left( \int d\vec{k} g(\vec{k}) a^{\dag}(\vec{k}) \right).
\label{eqn:trasf_sq}
\end{equation}
%%%%%%%%%%%%%%%%%%%%%%%%%%%%%%%%%%%%%%%%%%%
Clearly, the squeeze operator given above is similar to two mode squeeze operator 
in quantum mechanics.
%%%%%%%%%%%%%%%%%%%%%%%%%%%%%%%%%%%%%%%%%%%%%%%%%%%%%%%%%%%%%%%%%%%%%%%%%%%
Other examples of squeeze operator are shown in appendix~\ref{app:sqopexample}.

%%%%%%%%%%%%%%%%%%%%%%%%%%%%%%%%%%%%%
\section{Description of a Domain}
\label{sec:domain}
In this section, we attempt to describe a domain by a squeezed state.
For instance, a domain will be formed in a high energy heavy ion collision. 
In such a situation, the possibility of the formation of a squeezed state was pointed out
\cite{Amado, Hiro-oka_PLB, Hiro-oka_PLB_E, Hiro-oka_PRC}.
A certain squeezed state is constructed by using the squeezed operator defined 
in the previous section, and 
momentum distribution, chaoticity and correlation length, are calculated.

\subsection{The State of a Domain}
A coherent state which describes a domain \cite{our} is constructed 
with the displacement operator, eq.~\eqref{def:disp}:  
\begin{equation}
|{\rm co}\rangle = D(\alpha) |0\rangle , 
\end{equation}
where the vacuum state $|0\rangle$ is defined by $a(k)|0\rangle = 0$. 
A squeezed state is constructed with the squeeze operator defined by eq.~\eqref{def:contS}:
%%%%%%%%%%%%%%% Equation %%%%%%%%%%%%%%%
\begin{equation}
|{\rm sq}\rangle =  D(\alpha) S(G) |0 \rangle.
\label{eqn:sqst}
\end{equation}
%%%%%%%%%%%%%%% Equation END %%%%%%%%%%%%%%%
The squeezed state is called 'squeezed vacuum' when $\alpha(\vec{k})$ is equal to 0.

We attempt to obtain the state of a domain: 
the spatial distribution of the condensate is spherical and Gaussian,
%%%%%%%%%%%%%%
\begin{equation}
\langle \phi(0,\vec{x}) \rangle = Q \exp \left( - \frac{|\vec{x}|^{2}}{2R^{2}} \right) ,
\label{eqn:phi_dist}
\end{equation}
%%%%%%%%%%%%
where $\langle \cdots \rangle$ represents the expectation value.
The quantity $Q$ is the value of the condensates at the center of the domain and 
$R$ represents the size of the domain. 
%%%%%%%%%%%%
The expectation value of $\dot{\phi}(0,\vec{x})$, which is the time derivative of $\phi$, 
should be approximately zero for the domain to survive for a long time.
These requirements lead to 
\begin{equation} 
\alpha(\vec{k}) = \sqrt{\frac{\omega(\vec{k})}{2}} Q R^3  \exp\left( -\vec{k}^2 R^2/2 \right) .
\end{equation} 

%%%%%%%%%%%%%%%%%%
A squeezed states is related to the fluctuation which is measured by the following quantity:
%%%%%%%%%%%%%%%%%%%%%%%%%%%
\begin{equation}
D(t,\vec{x})
= \langle :\left( \phi^{2} (t,\vec{x}) \right): \rangle  - \langle \phi(t,\vec{x}) \rangle ^{2},
\label{eqn:fluc}
\end{equation}
%%%%%%%%%%%%%%%%%%%%%%%%%%%
where the symbol $:O(a,a^{\dag}):$ represents the normal ordering for the operators, 
$a(\vec{k})$ and $a^{\dag}(\vec{k})$.
%%%%%%%%%%%%%%%%%%%%%%%%%%%%%%%%%%%%%%%%%%%%%%%%%%%%%%%%%

We attempt to determine the squeeze operator. 
The squeeze operator introduced by Eq.~\eqref{def:contS} is too general.
We use the decoupling type of $G$ shown in the previous section.
The transformation of annihilation operator $a(\vec{k})$ is already given 
by Eq.~(\ref{eqn:trasf_sq}) .

The distribution of the fluctuation is assumed as follows:
%%%%%%%%%%%%%%%%%%%%%%%%%%% 
\begin{equation}
D(0,\vec{x}) = P \exp \left( - \frac{|\vec{x}|^{2}}{2H^{2}} \right),
\label{eqnflc-assump}
\end{equation}
%%%%%%%%%%%%%%%%%%%%%%%%%%%
where $P$ is the magnitude of squeezing at $|\vec{x}| = 0$ and 
$H$ represents the size of the squeezed region. 
If $P$ is equal to zero, the state is a coherent state.  
The quantity ${\displaystyle \left. dD(t,\vec{x})/dt \right|_{t=0}}$ 
should be nearly equal to zero 
in order that the squeezed region survives for a long time.

The equation for $g(\vec{k})$ is given by Fourier transformation to Eq.~\eqref{eqnflc-assump}:
\begin{align}
&\int \frac{d\vec{q}}{\sqrt{\omega(\vec{q}+\vec{p}/2) \ \omega(\vec{q}-\vec{p}/2)}}
\left\{ \frac{2 \sinh^2 L}{L} 
\ g(\vec{q}+\vec{p}/2) g^{*}(\vec{q}-\vec{p}/2) 
\right.
\nonumber \\
& \left.
- \frac{\cosh L \sinh L}{L}
\Big[
g(\vec{q}+\vec{p}/2) g(-(\vec{q}-\vec{p}/2))
+
g^{*}(\vec{q}-\vec{p}/2) g^{*}(-(\vec{q}+\vec{p}/2))
\Big]
\right\}
\nonumber \\
& = 
8PH^6
\int d\vec{q} 
\exp \left( - \left(\vec{q} + \frac{\vec{p}}{2}\right)^2 H^2 \right) 
\exp \left( - \left(\vec{q} - \frac{\vec{p}}{2}\right)^2 H^2 \right) 
,
\label{eqn:D_fourie}
\end{align}
where $L$ is defined by 
\begin{equation}
L := \int d^{3}\vec{k} g(\vec{k}) g^{*}(\vec{k}). 
\label{eqn:L}
\end{equation}
%%%%%%%%%%%%%%%%%%%%%%%%%%%%%%%%%%%%%%%%%%%%%%%%%%%%%%%%%%%%%%%%%%
The following functions satisfy Eq.~\eqref{eqn:D_fourie} clearly:
%%%%
\begin{subnumcases}{g(\vec{k}) = }
2H^{3}
\ \sqrt{\frac{-2P L }{\sinh( L ) \left( \cosh( L )-\sinh( L ) \right)}}
\ \sqrt{ \omega(\vec{k})} \ \exp \left( -k^{2}H^{2} \right) 
%%%%
\qquad P \leq 0  
\label{eqn:gr}
\\
%%%%%%%%%%%%%%%
2H^{3}i 
\ \sqrt{\frac{2P L }{\sinh( L ) \left( \cosh( L )+\sinh( L ) \right) }}
\ \sqrt{ \omega(\vec{k})} \ \exp \left( -k^{2}H^{2} \right) 
%%%%  
\qquad
P \geq 0
\label{eqn:gi}
,
\end{subnumcases}
where $\omega(\vec{k})$ is the free particle energy of the field $\phi$.

%%%%%%%%%%%%%%%%%%%%%%%%%%%
These functions $g$ have the constraint, Eq.~\eqref{eqn:L}.
The quantity $L$ is obtained by substituting Eq.~\eqref{eqn:gr} into Eq.~\eqref{eqn:L}
for $P \leq 0$
and Eq.~\eqref{eqn:gi} into Eq.~\eqref{eqn:L} for $P \geq 0$:
\begin{equation}
L= \frac{\mathrm{sgn}(P)}{2} \ln \left[1+64\pi PH^6 \chi(H) \right],
\end{equation}
where $\mathrm{sgn}(P)$ is 1 for $P \ge 0$ and $-1$ for $P < 0$,
and the function $\chi(H)$ is defined by 
\begin{equation}
\chi(H) := \int_{0}^{\infty} \ dk \ k^2 \omega(k) \exp(-2 k^2 H^2).
\end{equation}
We note that $P$ has the lower bound, 
because $L$ is positive from Eq.~\eqref{eqn:L}: 
\begin{equation}
-\frac{1}{64 \pi H^6 \chi(H)} < P .
\end{equation}
%%%%%%%%%%%%%%%%%%%%%%%%%%%%%%%%%%
It is easily shown that 
the functions, \eqref{eqn:gr} and \eqref{eqn:gi}, satisfy 
${\displaystyle \left. dD(t,\vec{x})/dt \right|_{t=0} = 0}$.
The state with Eqs.~\eqref{eqn:gr} or \eqref{eqn:gi} is surely the state of 
the long-lived domain with the distributions.

%%%%%%%%%%%%%%%%%%%%%%%%
\subsection{Observables}
In this subsection, 
momentum distribution, 
correlation function and correlation length are calculated and discussed. 

The momentum distribution in a spherical system is characterized by the absolute value of 
the momentum $k = |\vec{k}|$. That is, 
\begin{equation}
\frac{dN}{dk} = 
4\pi k^{2} 
\langle {\rm sq}| n(\vec{k}) |{\rm sq} \rangle, 
\end{equation}
where $n(\vec{k}) = a^{\dag}(\vec{k}) a(\vec{k})$.
This is easily evaluated by using eq. (\ref{eqn:trasf_sq}):
\begin{equation}
\frac{dN}{dk} = 
4\pi k^{2} \left[ 
\left| \alpha(\vec{k}) \right|^2 + \frac{\sinh^2 L}{L} \left| g(\vec{k}) \right|^2
\right]
\label{eqn:momentum_distribution}
\end{equation}
%%%%%%%%%%%%%%%%%%%%%%%%%%%%%%
Chaoticity $\lambda(\vec{k})$ is defined by
\begin{equation} 
\lambda(\vec{k}) := 
\frac{\Big\langle : \left( n(\vec{k}) \right)^2 : \Big\rangle }{\Big\langle n(\vec{k}) \Big\rangle^2} - 1 .
\label{eqn:chao}
\end{equation}
In the present case, the chaoticity is given by 
\begin{align}
\lambda(\vec{k}) 
= & \frac{1}{L^2 \langle n(\vec{k}) \rangle^{2}} \left\{
  \sinh^2 L \left(\cosh^2 L + \sinh^2 L\right)\left|g(\vec{k})\right|^4
+ 2L \sinh^2 \left|\alpha(\vec{k})\right|^2 \left|g(\vec{k})\right|^2
\right. 
\nonumber \\
& \left.  
- L \cosh L \sinh L \left[
\left( \alpha(\vec{k})\right)^2 \left( g^*(\vec{k}) \right)^2
+
\left( \alpha^*(\vec{k})\right)^2 \left( g(\vec{k}) \right)^2
 \right]
\right\} .
\end{align}
%%%%%%%%%%%%%%%%%%%%%%%%%%%%%

The correlation function 
$\left. \left\langle {\mathrm{sq}} \right| :\phi(t,\vec{x}) \phi(t,\vec{y}) :\left|{\mathrm{sq}} \right\rangle \right|_{t=0}$
is calculated by using the functions $\alpha(\vec{k})$ and $g(\vec{k})$:
\begin{equation}
\left\langle {\mathrm{sq}} \right| :\phi(0,\vec{x}) \phi(0,\vec{y}) :\left|{\mathrm{sq}} \right\rangle 
= Q^2\exp \left( -\frac{(|\vec{x}|^2+|\vec{y}|^2)}{2R^2} \right) 
 + P \exp \left( -\frac{(|\vec{x}|^2+|\vec{y}|^2)}{4H^2} \right),
\end{equation}
where the first term is the contribution of the coherent part 
and the second term is the contribution of the squeezed part.
We define the correlation length $\xi$ by replacing $|\vec{x}|^2+|\vec{y}|^2$ by $\xi^2$.
The length $\xi_s$ that is defined by $\xi/R$ satisfies
\begin{equation}
\exp \left( - \frac{1}{2} \ \xi_s^2 \right) 
+ \left( \frac{P}{Q^2} \right) \exp \left( - \left( \frac{R^2}{4H^2} \right) \ \xi_s^2 \right) 
= \left(1 + \frac{P}{Q^2}\right) \exp \left(-1\right) . 
\label{eq:xis}
\end{equation}
Clearly, $\xi$ is $\sqrt{2} R$ for the coherent state
and $2H$ for the squeezed vacuum. 
The length $\xi_s$ is $\sqrt{2}$ and is independent of $P/Q^2$, when $R$ equals $\sqrt{2} H$.
For strong squeezing ($P \gg Q^2$), $\xi_s$ is approximately equal to $2H/R$. 

%%%%%%%%%%%%%%%%%%%%%%%%%%%%%%%%%%%%%%%%%%%%%%%%%%%%%%%%%%%%%%%%%%%%%%%%%%%%%%
\subsection{Numerical Examples}
\label{sec:numerical}
In this subsection, 
the expectation values of physical quantities by the squeezed state are shown numerically.
We define dimensionless quantities:
$\vec{\mu} := R \vec{k}$, $\rho := P/Q^2$, $\sigma := QR$, $\eta := H/R$
and $m_R := m R$, where $m$ is the mass of the field $\phi$.
We calculate the momentum distribution, $dN/d\mu$, and the chaoticity $\lambda$
for various values of the parameters, where $\mu$ is the absolute value of $\vec{\mu}$.

\begin{figure}
\begin{center}
\includegraphics[width=0.5\textwidth]{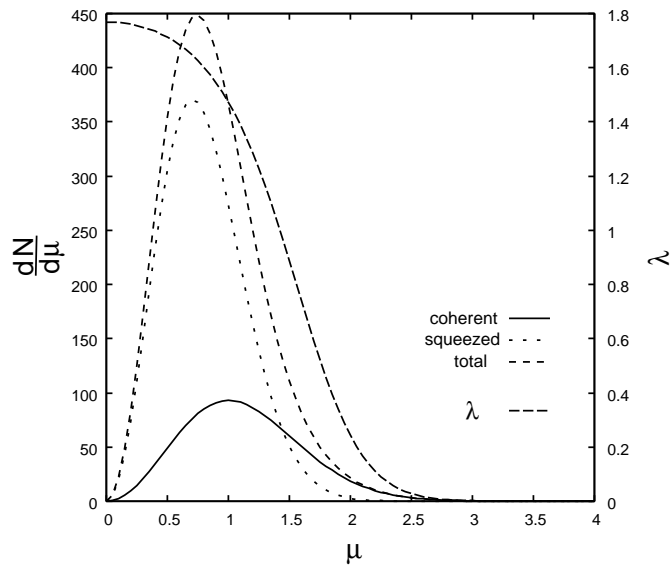}
\end{center}
\caption{
Momentum distribution $dN/d\mu$ and chaoticity $\lambda$ for 
$\rho=1.0$, $\eta=1.0$, $\sigma=2.0$ and $m_R=10.0$ .
The solid line is the momentum distribution caused by the coherent part, 
the dotted line is that caused by the squeezed part,  
the dashed line is the total momentum distribution,  
and the long-dashed line is the chaoticity. 
}
\label{fig:momentum_distribution:basic}
\end{figure}

%%%%%%%%%%%%%%%%%%%%%%%
An example is shown in Figure \ref{fig:momentum_distribution:basic}.
This figure shows the momentum distribution and the chaoticity
for $\rho=1.0$, $\eta=1.0$, $\sigma=2.0$ and $m_R=10.0$.
The momentum distribution is divided into two parts, coherent and squeezed parts, 
in Eq.~\eqref{eqn:momentum_distribution}.
The typical momentum is approximately the inverse of the domain size: $R^{-1}$ or $H^{-1}$. 
The quantity $dN/d\mu$ caused by the coherent part 
is larger than that caused by squeezed part for large $\mu$.
Therefore the chaoticity approaches zero as $\mu$ approaches infinity.

%%%%%%%%%%%%%
%%%%%%%%%%

\begin{figure}
\begin{center}
\begin{tabular}{ccc}
\includegraphics[width=0.33\textwidth]{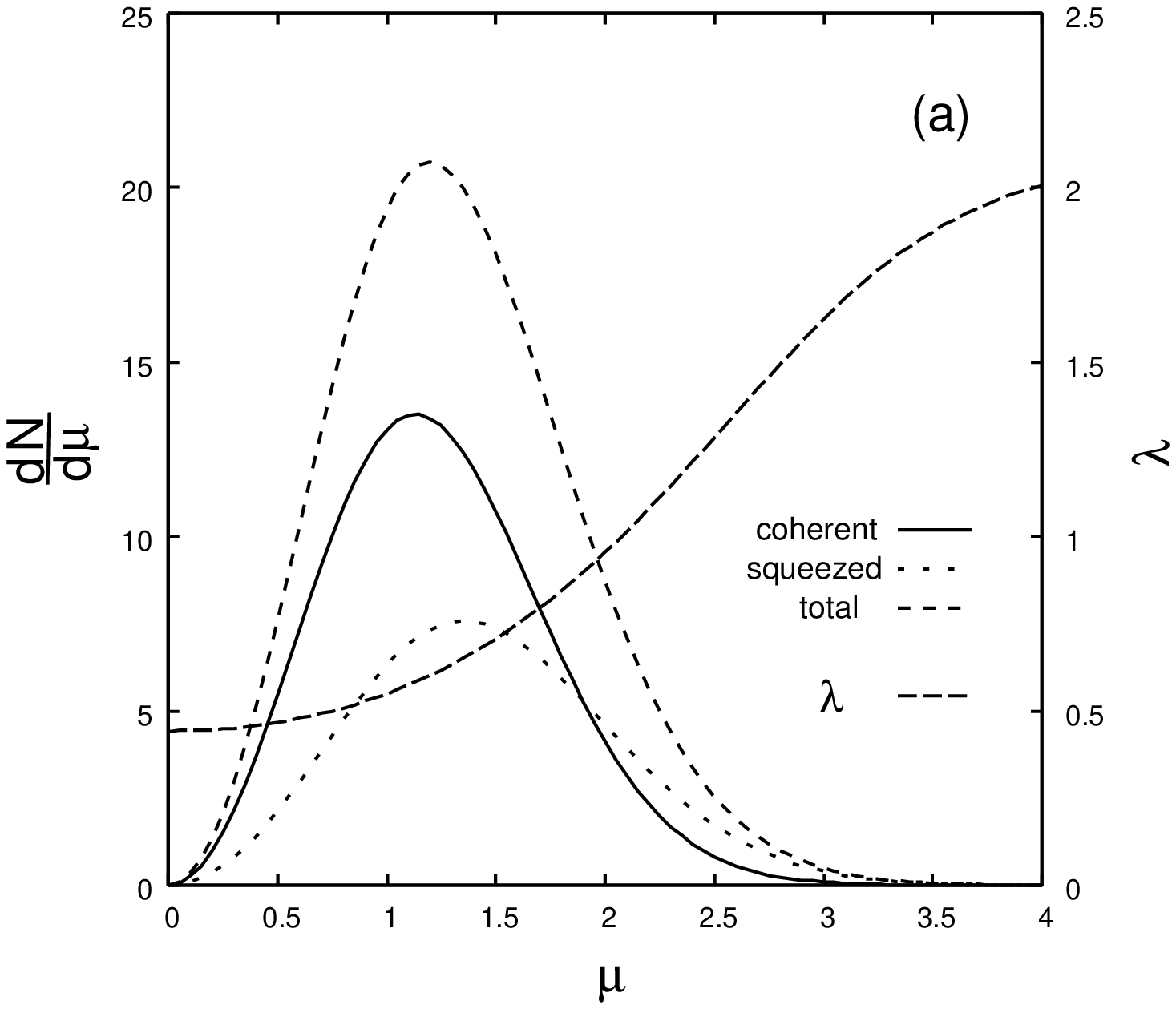}
&
\includegraphics[width=0.33\textwidth]{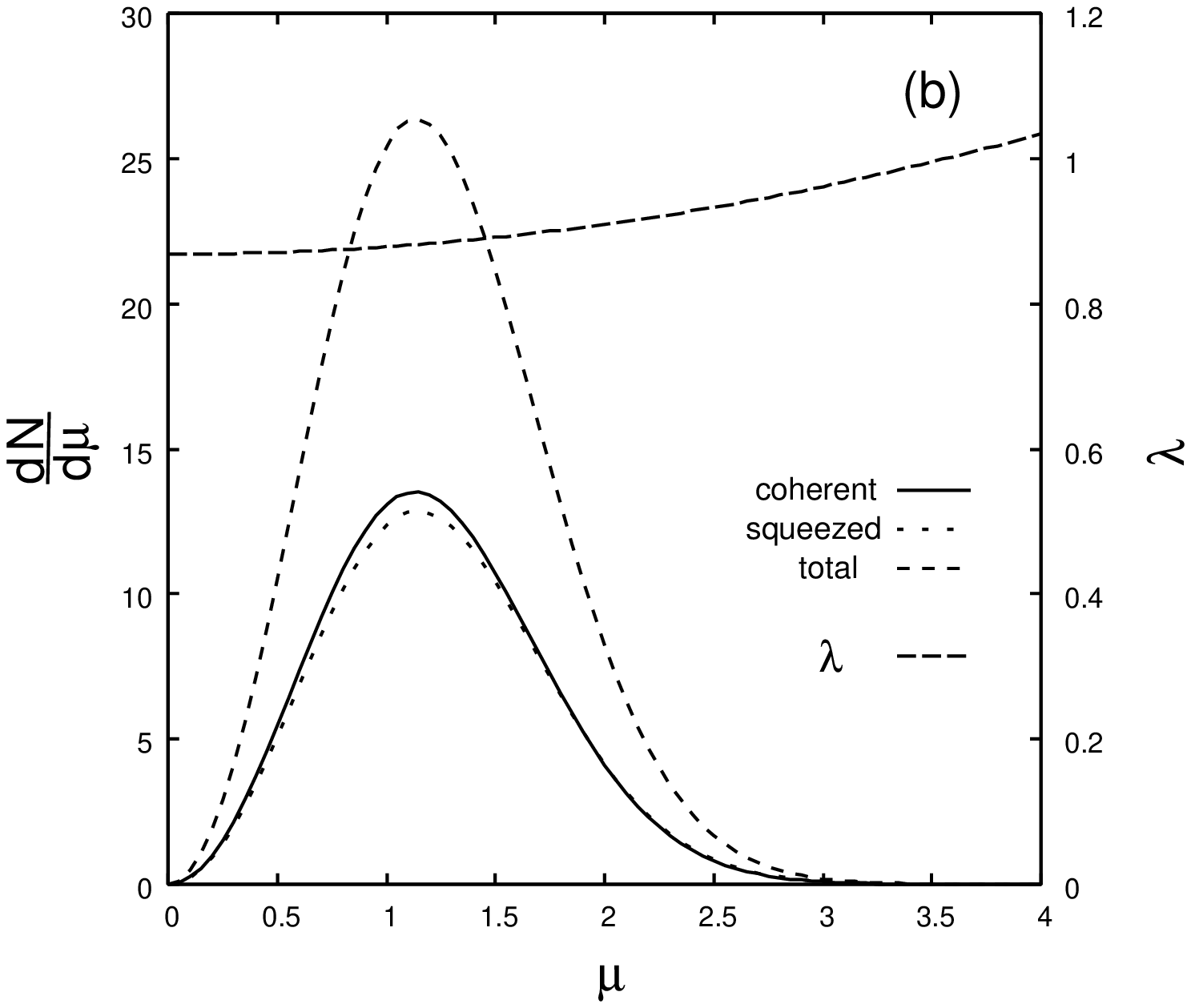}
&
\includegraphics[width=0.33\textwidth]{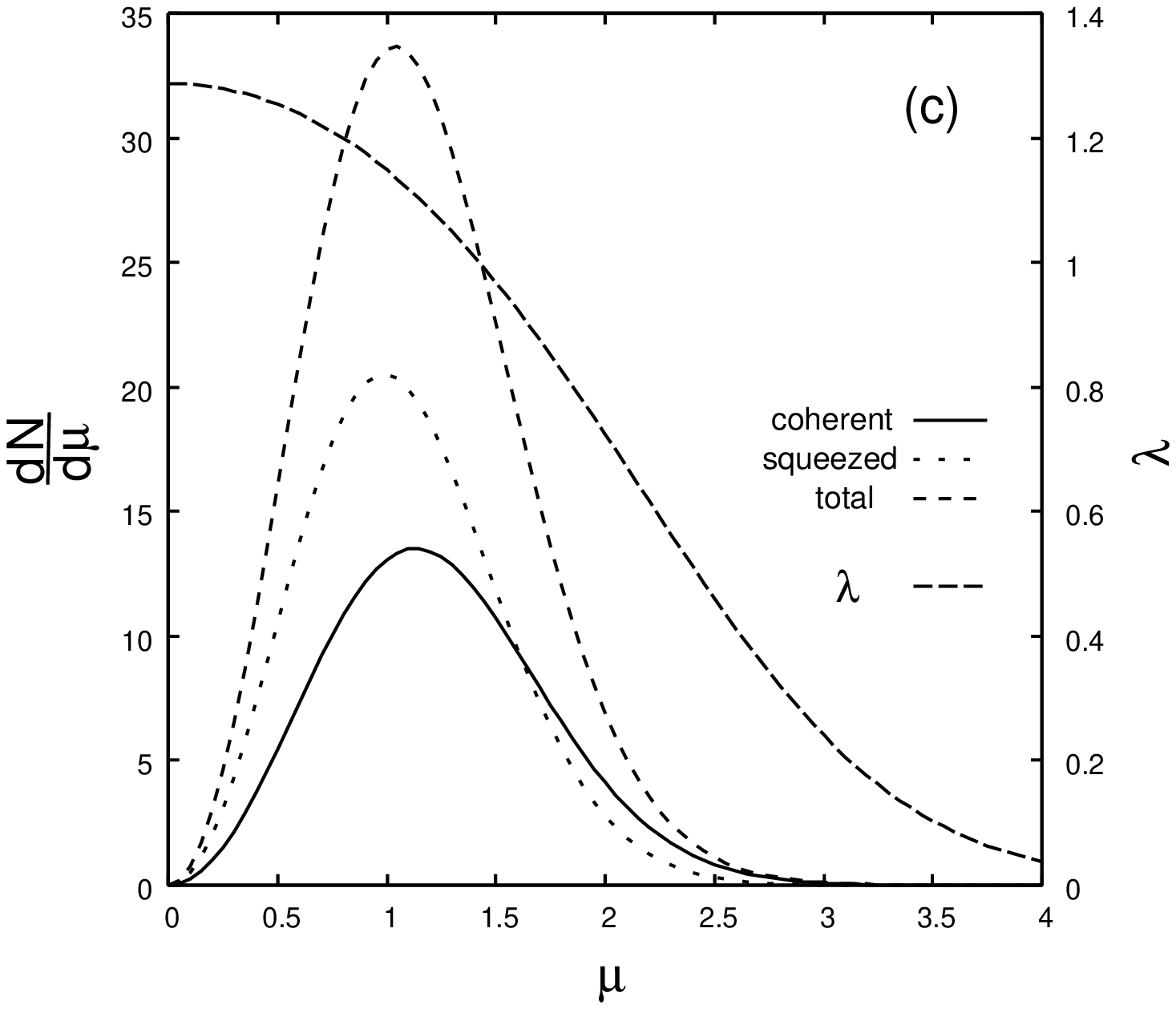}
\end{tabular}
\end{center}
\caption{Momentum distribution $dN/d\mu$ and chaoticity $\lambda$ 
for (a) $\eta = 0.6$, (b) $\eta = 0.7$ and (c) $\eta = 0.8$.
The other parameters are $\rho=1.0$, $\sigma=2.0$ and  $m_R=1.0$. 
The solid line is the momentum distribution caused by the coherent part, 
the dotted line is that caused by the squeezed part,  
the dashed line is the total momentum distribution,  
and 
the long-dashed line is the chaoticity. 
}
\label{fig:momentum_distribution}
\end{figure}

%%%%%%%%%%%%%%%%%%%%%%%
Figure~\ref{fig:momentum_distribution} shows the momentum distribution and the chaoticity
for (a) $\eta = 0.6$, (b) $\eta = 0.7$ and (c) $\eta = 0.8$.
The other parameters are $\rho=1.0$, $\sigma=2.0$ and  $m_R=1.0$. 
As shown in Fig.~\ref{fig:momentum_distribution}, 
the chaoticity depends strongly on $\eta$ in the vicinity of $\eta=0.7$.
%%%%%%%
In Fig.~\ref{fig:momentum_distribution}(a), 
the quantity $dN/d\mu$ caused by the squeezed part is larger than that 
caused by coherent part for large $\mu$. 
Therefore the chaoticity does not approach zero as $\mu$ approach infinity.
In Fig.~\ref{fig:momentum_distribution}(b), 
the quantity $dN/d\mu$ caused by the squeezed part is approximately equal to 
that caused by coherent part in the wide range of $\mu$.
Therefore the chaoticity is approximately constant.
In Fig.~\ref{fig:momentum_distribution}(c), 
the quantity $dN/d\mu$ caused by coherent part is larger than 
that caused by the squeezed part for large $\mu$.
The chaoticity approaches zero as $\mu$ approaches infinity.
These figures indicate that the chaoticity notifies us of the size of the squeezing region compared 
with the size of coherent region.
The chaoticity will be a signal of the squeezing, while the momentum distribution will not.

%%%%%%%%%%
%%%%%%%%%%

\begin{figure}
\begin{center}
\includegraphics[width=0.45\textwidth]{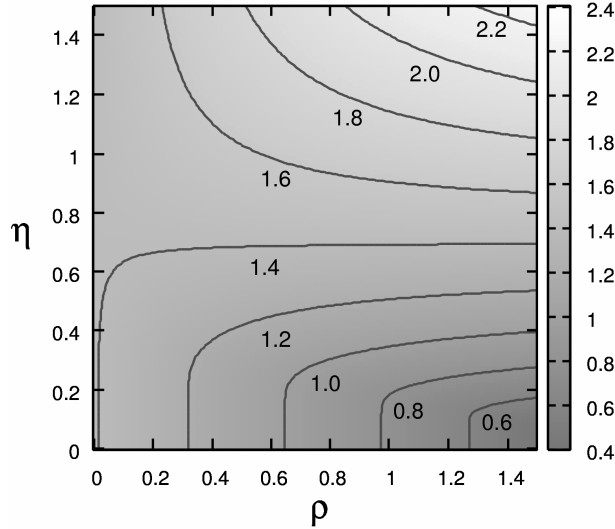}
\end{center}
\caption{Length $\xi_s$ on the $\rho$-$\eta$ plane.}
\label{fig:density_plot_of_xis}
\end{figure}

Figure \ref{fig:density_plot_of_xis} shows the density plot of the correlation length $\xi_s$ 
with contour lines.
The value $\xi_s$ is given by Eq.~\eqref{eq:xis}. 
From the figure,
the $\eta$ dependence is strong for large $\rho$ and weak for small $\rho$.
The correlation length is determined by $\eta$ in the case of strong squeezing ($\rho \gg 1$).
%%%%%%%%%%%%%%%%%%%%%%%%%%%%%%%%%%%%%%%%%%%%%%%%%%%%%%%%%%%%%%%%%%%%%%%%%%%%%%%%

\begin{figure}
\begin{center}
\includegraphics[width=0.45\textwidth]{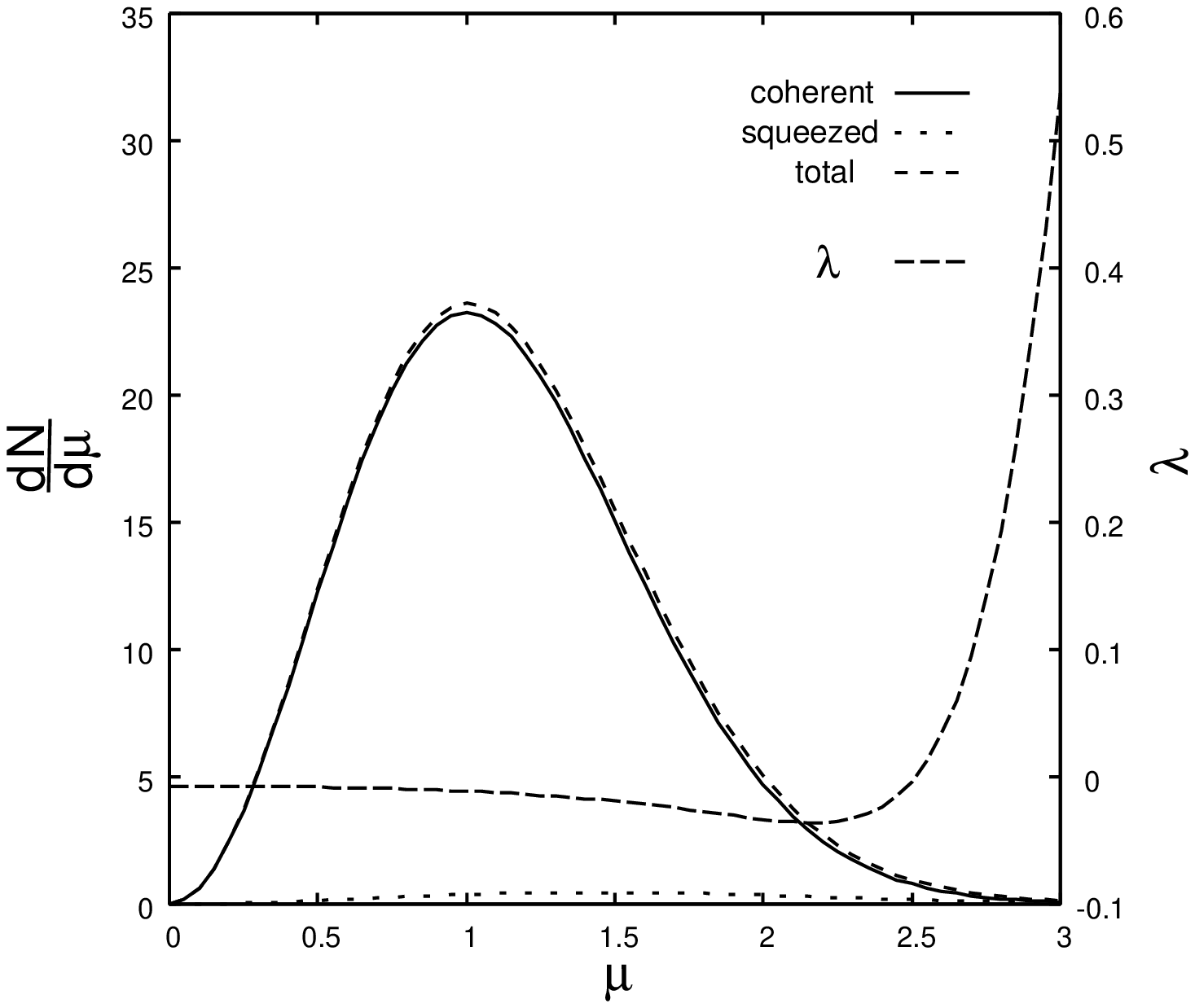}
\end{center}
\caption{
Momentum distribution $dN/d\mu$ and chaoticity $\lambda$ for 
$\rho=-0.02$, $\eta=0.50$, $\sigma=1.00$ and $m_R=10.00$.
}
\label{fig:negativeP}
\end{figure}

%%%%%%%%%%%%%%%%%%%%%%%%%
The quantities for negative $\rho$ are calculated in the same manner. 
Figure~\ref{fig:negativeP} shows the momentum distribution and the chaoticity 
for $\rho=-0.02$, $\eta=0.50$, $\sigma=1.00$ and $m_R=10.00$.
Clearly, the chaoticity defined by Eq.~\eqref{eqn:chao} is negative in the region 
where the distribution caused by squeezed part is not zero. 

%%%%%%%%%%%%%%%%%%%%%%%%%
Finally, we deal with a pion field to give an concrete example. 
We assume that the distribution of the pion condensate and of the fluctuation are Gaussian.
In these numerical calculations, 
we set the parameters: $Q=f_{\pi}$, $P = f_{\pi}^2$ and $m=135$MeV, %/$\mathrm{c}^2$,
where $f_{\pi}$ is pion decay constant.
This constant is taken to be 90MeV in the present calculations.
The remaining parameters are set to (a) $R=H=5$fm, and (b) $R=5$fm and $H=3$fm.

\begin{figure}
\begin{center}
\begin{tabular}{cc}
\includegraphics[width=0.45\textwidth]{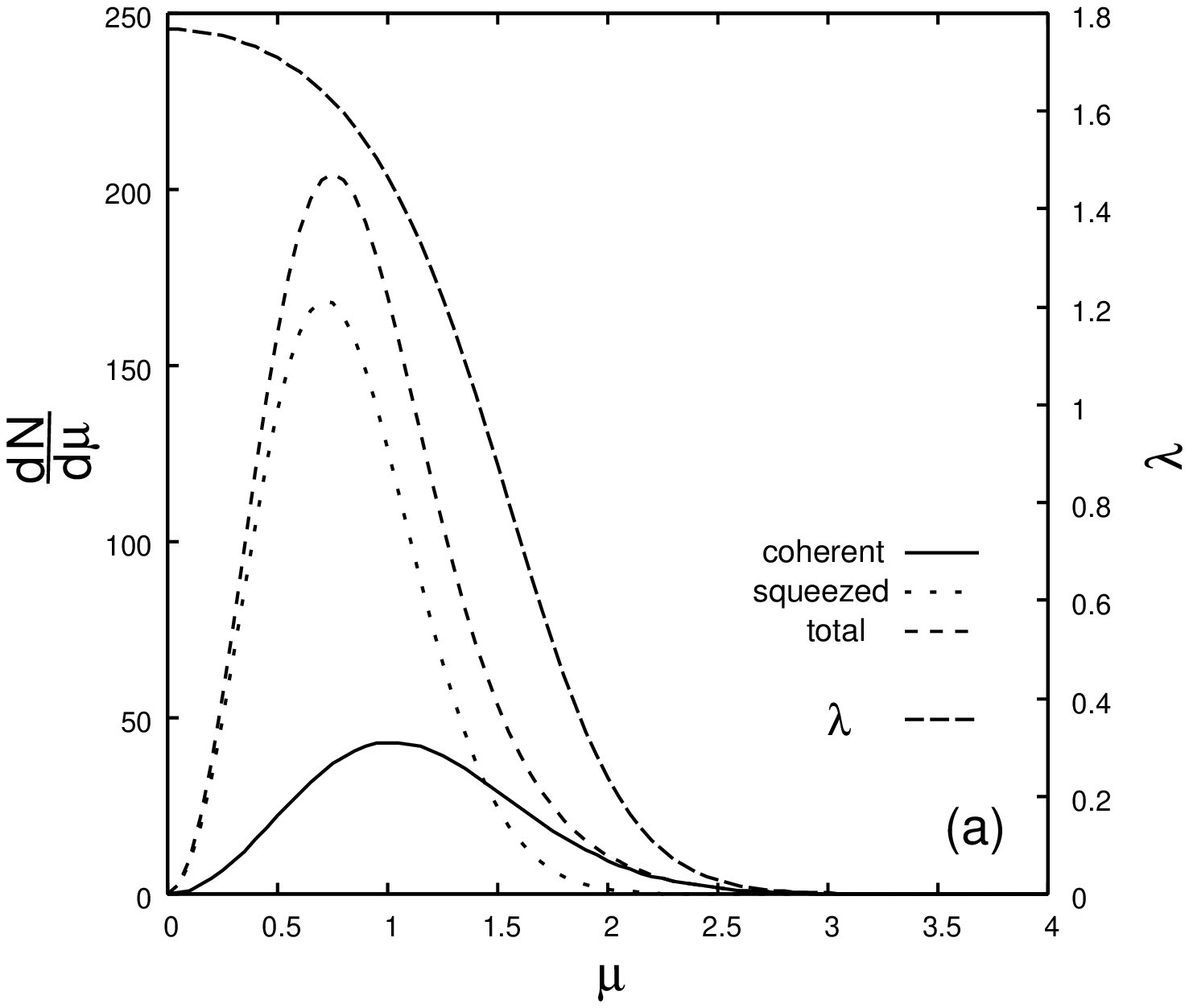}
&
\includegraphics[width=0.45\textwidth]{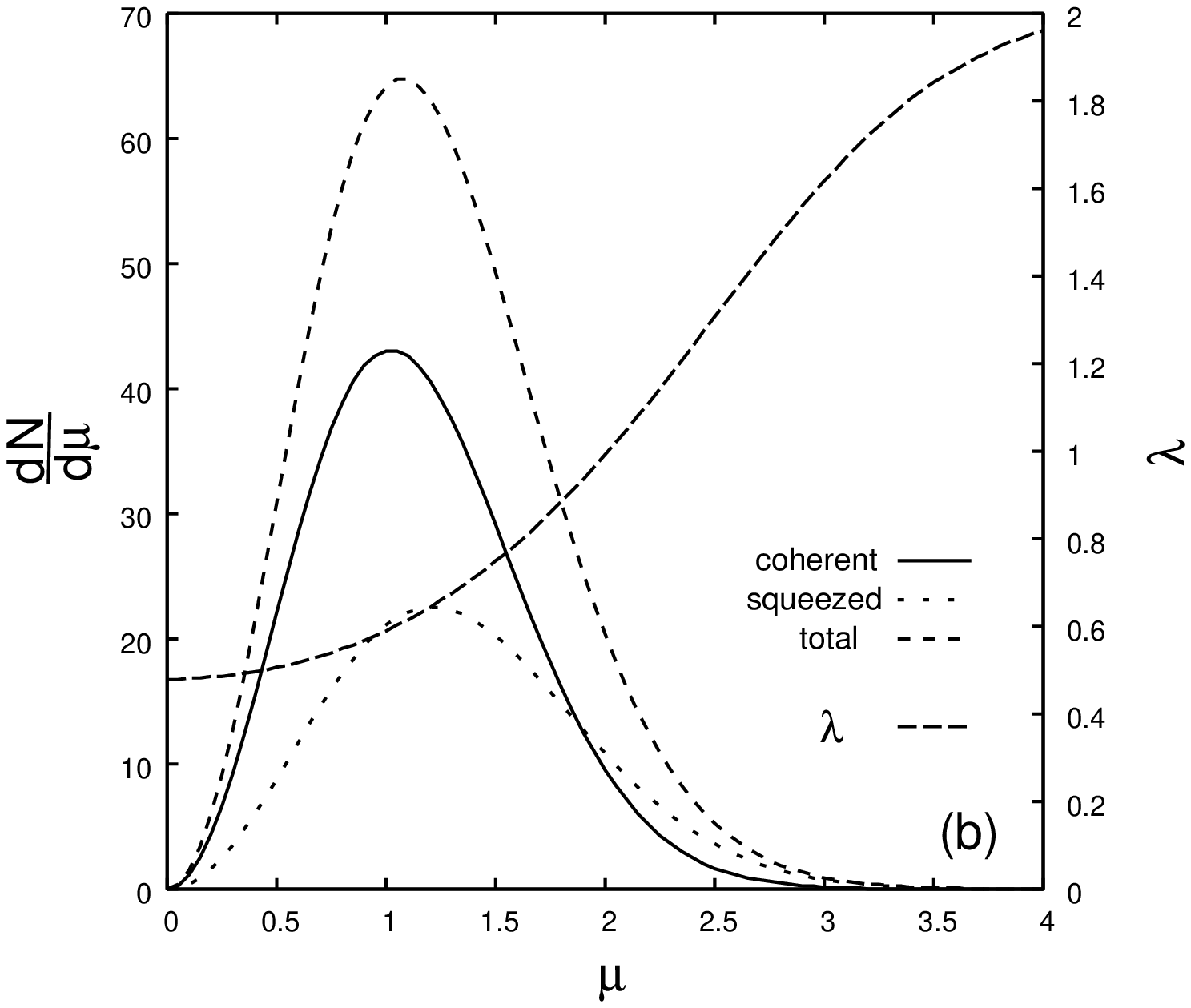}
\end{tabular}
\end{center}
\caption{Momentum distribution emitted instantaneously from a domain of pions.
The parameters, $R$ and $H$, are (a) $R=H=5$fm and (b) $R=5$fm, $H=3$fm.
The other parameters are $Q=f_{\pi}=90{\mathrm{MeV}}$, $P=f_{\pi}^2$ and $m=135 \mathrm{MeV}$.
The dimensionless variable $\mu$ describes the magnitude of the momentum 
and $\lambda$ is chaoticity.
The solid line is the momentum distribution caused by the coherent part, 
the dotted line is that by the squeezed part,  
the dashed line is the total momentum distribution,  
and 
the long-dashed line is the chaoticity. 
}
\label{fig:mom_pion}
\end{figure}

Figure~\ref{fig:mom_pion} shows the momentum distribution of pions
when the domain decays instantaneously.
The distribution for the squeezed  part reflects the size of the squeezed region,
the parameter $H$.  
%%%%%%%%%%%%%%%%%%%%%
The distributions in the case of $H=5$fm are depicted in Fig.~\ref{fig:mom_pion}(a)
and those in the case of $H=3$fm in Fig.~\ref{fig:mom_pion}(b).
%%%%%%%%%%%%%%%%%%%%%
Figure~\ref{fig:mom_pion}(a) is resemble to Fig.~\ref{fig:momentum_distribution}(c),
because $\eta$ is equal to 1.
On the contrary,
Fig.~\ref{fig:mom_pion}(b) is resemble to Fig.~\ref{fig:momentum_distribution}(a),
because $\eta$ is equal to 0.6.
The chaoticity approaches zero in the case of $H=5$fm, 
and does not equal zero for large $\mu$ in the case of $H=3$fm.
The chaoticity should be a signal of squeezing when the state of the pion field is squeezed
in high energy collisions.
%%%%%%%%%%%%%%%%%%%%%

%%%%%%%%%%%%%%%%%%%%%%%%%%%%%%%%%%%%%
\section{Conclusion and Discussion}
\label{sec:conclusion}
In this paper, we attempted to describe the quantum state of a domain in quantum field theory.  
We used the extended squeeze operator
and assumed that the distributions of the condensate and of the fluctuation are Gaussian.
The state of a domain was given explicitly by using displacement and squeezing operators.
The momentum distribution, the chaoticity and the correlation length were calculated. 
As an example, we dealt with a pion field.

%%%%%%%%%%%%
The order of the momentum distribution is clearly given by the inverse of the domain size: $R^{-1}$ or $H^{-1}$. 
The behavior of the chaoticity changes markedly as a function of $\eta$ that is defined by $H/R$.
Therefore we can evaluate $\eta$ by the chaoticity:
the quantity $\eta$ is less than, equal to, or greater than about $0.7$, 
because the behavior changes in the vicinity of $\eta = 0.7$.
%%%%%%
This behavior is explained by the behaviors of the function that describes the condensate, $\alpha(\vec{k})$, 
and of the function that describes the fluctuation, $g(\vec{k})$.   
The function $\alpha(\vec{k})$ behaves roughly $\exp(-k^2 R^2/2)$ and
$g(\vec{k})$ behaves roughly $\exp(-k^2 H^2)$.
Therefore 
the behavior of $\alpha(\vec{k})$ is similar to that of $g(\vec{k})$ 
when $R$ is close to $\sqrt{2} H$:
$\eta = H/R \sim 1/\sqrt{2}\sim 0.7$. 

%%%%%%
The correlation function is not translational invariant in the present system.
Therefore we defined the correlation length that is measured from the center of the domain. 
The $\eta$ dependence of this correlation length is weak for small $\rho$ 
and is strong for large $\rho$, where $\rho$ is $P/Q^2$.
The correlation length $\xi$ is $\sqrt{2} R$ and is independent of $\rho$ when $\eta = 1/\sqrt{2}$.
The correlation length is long 
when the squeezed region is large (large $\eta$) and the strength of the squeezing is strong (large $\rho$).

%%%%%%%%%% 
In summary, 
the quantum state of a domain are constructed by using the extended squeeze operator, 
when the distributions of the condensate and of the fluctuation are Gaussian. 
Momentum distribution, chaoticity are calculated by using the state.
Inversely, the approximate size of the domain is inferred by the momentum distribution.
The ratio of the size of the squeezed region to that of the coherent region is inferred by the chaoticity. 

We hope that this work is helpful to describe domains in quantum field theories.

%%%%%%%%%%%%%%%%%%%%%%%%%%%%%%%%%%%%%%%%%%%%%%%%%%%%%%%%%%%%%%%%%%%%%%%%%%%%%%%%%%%%%%%%%%%%%%%%%%%%%%%%%%%%%%%
%%%% To make bibliography
%%%\bibliographystyle{apsrev4-1}
%%%\bibliography{squeeze.bib}

\begin{thebibliography}{18}%
\makeatletter
\providecommand \@ifxundefined [1]{%
 \@ifx{#1\undefined}
}%
\providecommand \@ifnum [1]{%
 \ifnum #1\expandafter \@firstoftwo
 \else \expandafter \@secondoftwo
 \fi
}%
\providecommand \@ifx [1]{%
 \ifx #1\expandafter \@firstoftwo
 \else \expandafter \@secondoftwo
 \fi
}%
\providecommand \natexlab [1]{#1}%
\providecommand \enquote  [1]{``#1''}%
\providecommand \bibnamefont  [1]{#1}%
\providecommand \bibfnamefont [1]{#1}%
\providecommand \citenamefont [1]{#1}%
\providecommand \href@noop [0]{\@secondoftwo}%
\providecommand \href [0]{\begingroup \@sanitize@url \@href}%
\providecommand \@href[1]{\@@startlink{#1}\@@href}%
\providecommand \@@href[1]{\endgroup#1\@@endlink}%
\providecommand \@sanitize@url [0]{\catcode `\\12\catcode `\$12\catcode
  `\&12\catcode `\#12\catcode `\^12\catcode `\_12\catcode `\%12\relax}%
\providecommand \@@startlink[1]{}%
\providecommand \@@endlink[0]{}%
\providecommand \url  [0]{\begingroup\@sanitize@url \@url }%
\providecommand \@url [1]{\endgroup\@href {#1}{\urlprefix }}%
\providecommand \urlprefix  [0]{URL }%
\providecommand \Eprint [0]{\href }%
\providecommand \doibase [0]{http://dx.doi.org/}%
\providecommand \selectlanguage [0]{\@gobble}%
\providecommand \bibinfo  [0]{\@secondoftwo}%
\providecommand \bibfield  [0]{\@secondoftwo}%
\providecommand \translation [1]{[#1]}%
\providecommand \BibitemOpen [0]{}%
\providecommand \bibitemStop [0]{}%
\providecommand \bibitemNoStop [0]{.\EOS\space}%
\providecommand \EOS [0]{\spacefactor3000\relax}%
\providecommand \BibitemShut  [1]{\csname bibitem#1\endcsname}%
\let\auto@bib@innerbib\@empty
%</preamble>
\bibitem [{\citenamefont {{D.~F.~Walls and G.~J.~Milburn}}(1994)}]{book:squ}%
  \BibitemOpen
  \bibfield  {author} {\bibinfo {author} {\bibnamefont {{D.~F.~Walls and
  G.~J.~Milburn}}},\ }\href@noop {} {\emph {\bibinfo {title} {Quantum
  Optics}}}\ (\bibinfo  {publisher} {Springer},\ \bibinfo {year}
  {1994})\BibitemShut {NoStop}%
\bibitem [{\citenamefont {Yuen}(1976)}]{Yuen}%
  \BibitemOpen
  \bibfield  {author} {\bibinfo {author} {\bibfnamefont {H.~P.}\ \bibnamefont
  {Yuen}},\ }\href@noop {} {\bibfield  {journal} {\bibinfo  {journal} {Phys.
  Rev. A}\ }\textbf {\bibinfo {volume} {13}},\ \bibinfo {pages} {2226}
  (\bibinfo {year} {1976})}\BibitemShut {NoStop}%
\bibitem [{\citenamefont {Walls}(1983)}]{Walls}%
  \BibitemOpen
  \bibfield  {author} {\bibinfo {author} {\bibfnamefont {D.~F.}\ \bibnamefont
  {Walls}},\ }\href@noop {} {\bibfield  {journal} {\bibinfo  {journal}
  {Nature}\ }\textbf {\bibinfo {volume} {306}},\ \bibinfo {pages} {141}
  (\bibinfo {year} {1983})}\BibitemShut {NoStop}%
\bibitem [{\citenamefont {Parker}(1969)}]{production_Parker}%
  \BibitemOpen
  \bibfield  {author} {\bibinfo {author} {\bibfnamefont {L.}~\bibnamefont
  {Parker}},\ }\href@noop {} {\bibfield  {journal} {\bibinfo  {journal} {Phys.
  Rev.}\ }\textbf {\bibinfo {volume} {183}},\ \bibinfo {pages} {1057} (\bibinfo
  {year} {1969})}\BibitemShut {NoStop}%
\bibitem [{\citenamefont {{D. Boyanovsky, H. J. de Vega and R.
  Holman}}(1994)}]{production_Boyanovsky}%
  \BibitemOpen
  \bibfield  {author} {\bibinfo {author} {\bibnamefont {{D. Boyanovsky, H. J.
  de Vega and R. Holman}}},\ }\href@noop {} {\bibfield  {journal} {\bibinfo
  {journal} {Phys. Rev. D}\ }\textbf {\bibinfo {volume} {49}},\ \bibinfo
  {pages} {2769} (\bibinfo {year} {1994})}\BibitemShut {NoStop}%
\bibitem [{\citenamefont {{H.~Mishra and A.~R.~Panda}}(1992)}]{EPSQS_H_Mishra}%
  \BibitemOpen
  \bibfield  {author} {\bibinfo {author} {\bibnamefont {{H.~Mishra and
  A.~R.~Panda}}},\ }\href@noop {} {\bibfield  {journal} {\bibinfo  {journal}
  {J. Phys. G}\ }\textbf {\bibinfo {volume} {18}},\ \bibinfo {pages} {1301}
  (\bibinfo {year} {1992})}\BibitemShut {NoStop}%
\bibitem [{\citenamefont {{A.~Mishra, P.~K.~Panda, S.~Schramm, J.~Reinhardt and
  W.~Greiner}}(1997)}]{EPSQS_A_Mishra_56}%
  \BibitemOpen
  \bibfield  {author} {\bibinfo {author} {\bibnamefont {{A.~Mishra,
  P.~K.~Panda, S.~Schramm, J.~Reinhardt and W.~Greiner}}},\ }\href@noop {}
  {\bibfield  {journal} {\bibinfo  {journal} {Phys. Rev. C}\ }\textbf {\bibinfo
  {volume} {56}},\ \bibinfo {pages} {1380} (\bibinfo {year}
  {1997})}\BibitemShut {NoStop}%
\bibitem [{\citenamefont {{A. Mishra and H.Mishra}}(1997)}]{EPSQS_A_Mishra_23}%
  \BibitemOpen
  \bibfield  {author} {\bibinfo {author} {\bibnamefont {{A. Mishra and
  H.Mishra}}},\ }\href@noop {} {\bibfield  {journal} {\bibinfo  {journal} {J.
  Phys. G}\ }\textbf {\bibinfo {volume} {23}},\ \bibinfo {pages} {143}
  (\bibinfo {year} {1997})}\BibitemShut {NoStop}%
\bibitem [{\citenamefont {{T.del R\'iogaztelurrutia and
  A.C.DAVIS}}(1990)}]{EPSQS_Tdel}%
  \BibitemOpen
  \bibfield  {author} {\bibinfo {author} {\bibnamefont {{T.del
  R\'iogaztelurrutia and A.C.DAVIS}}},\ }\href@noop {} {\bibfield  {journal}
  {\bibinfo  {journal} {Nucl. Phys. B}\ }\textbf {\bibinfo {volume} {347}},\
  \bibinfo {pages} {319} (\bibinfo {year} {1990})}\BibitemShut {NoStop}%
\bibitem [{\citenamefont {{M. Ishihara, M. Maruyama and F.
  Takagi}}(1998)}]{our}%
  \BibitemOpen
  \bibfield  {author} {\bibinfo {author} {\bibnamefont {{M. Ishihara, M.
  Maruyama and F. Takagi}}},\ }\href@noop {} {\bibfield  {journal} {\bibinfo
  {journal} {Phys. Rev. C}\ }\textbf {\bibinfo {volume} {57}},\ \bibinfo
  {pages} {1440} (\bibinfo {year} {1998})}\BibitemShut {NoStop}%
\bibitem [{\citenamefont {{M. Asakawa, T. Cs\"org\H o and M.
  Gyulassy}}(1999)}]{Asakawa_PRL}%
  \BibitemOpen
  \bibfield  {author} {\bibinfo {author} {\bibnamefont {{M. Asakawa, T.
  Cs\"org\H o and M. Gyulassy}}},\ }\href@noop {} {\bibfield  {journal}
  {\bibinfo  {journal} {Phys. Rev. Lett.}\ }\textbf {\bibinfo {volume} {83}},\
  \bibinfo {pages} {4013} (\bibinfo {year} {1999})}\BibitemShut {NoStop}%
\bibitem [{\citenamefont {{S.~S.~Padula}}\ \emph {et~al.}(2006)\citenamefont
  {{S.~S.~Padula}}, \citenamefont {{G.~Krein}}, \citenamefont
  {{T.~Cs\"org\H{o}}}, \citenamefont {{Y.~Hama}},\ and\ \citenamefont
  {{P.~K.~Panda}}}]{Sandra_BBC_2006}%
  \BibitemOpen
  \bibfield  {author} {\bibinfo {author} {\bibnamefont {{S.~S.~Padula}}},
  \bibinfo {author} {\bibnamefont {{G.~Krein}}}, \bibinfo {author}
  {\bibnamefont {{T.~Cs\"org\H{o}}}}, \bibinfo {author} {\bibnamefont
  {{Y.~Hama}}}, \ and\ \bibinfo {author} {\bibnamefont {{P.~K.~Panda}}},\
  }\href@noop {} {\bibfield  {journal} {\bibinfo  {journal} {Phy. Rev. C}\
  }\textbf {\bibinfo {volume} {73}},\ \bibinfo {pages} {044906} (\bibinfo
  {year} {2006})}\BibitemShut {NoStop}%
\bibitem [{\citenamefont {{T.~Cs\"org\H{o}}}\ and\ \citenamefont
  {{Sandra~S.~Padula}}(2007)}]{Csorgo_BBC_2007}%
  \BibitemOpen
  \bibfield  {author} {\bibinfo {author} {\bibnamefont {{T.~Cs\"org\H{o}}}}\
  and\ \bibinfo {author} {\bibnamefont {{Sandra~S.~Padula}}},\ }\href@noop {}
  {\bibfield  {journal} {\bibinfo  {journal} {Brazillian Journal of Physics}\
  }\textbf {\bibinfo {volume} {37}},\ \bibinfo {pages} {949} (\bibinfo {year}
  {2007})}\BibitemShut {NoStop}%
\bibitem [{\citenamefont {{I. V. Andreev and R. M. Weiner}}(1996)}]{contsqu}%
  \BibitemOpen
  \bibfield  {author} {\bibinfo {author} {\bibnamefont {{I. V. Andreev and R.
  M. Weiner}}},\ }\href@noop {} {\bibfield  {journal} {\bibinfo  {journal}
  {Phys. Lett. B}\ }\textbf {\bibinfo {volume} {373}},\ \bibinfo {pages} {159}
  (\bibinfo {year} {1996})}\BibitemShut {NoStop}%
\bibitem [{\citenamefont {{R. D. Amado and I. I. Kogan}}(1995)}]{Amado}%
  \BibitemOpen
  \bibfield  {author} {\bibinfo {author} {\bibnamefont {{R. D. Amado and I. I.
  Kogan}}},\ }\href@noop {} {\bibfield  {journal} {\bibinfo  {journal} {Phys.
  Rev. D}\ }\textbf {\bibinfo {volume} {51}},\ \bibinfo {pages} {190} (\bibinfo
  {year} {1995})}\BibitemShut {NoStop}%
\bibitem [{\citenamefont {{H. Hiro-Oka and H.
  Minakata}}(1998{\natexlab{a}})}]{Hiro-oka_PLB}%
  \BibitemOpen
  \bibfield  {author} {\bibinfo {author} {\bibnamefont {{H. Hiro-Oka and H.
  Minakata}}},\ }\href@noop {} {\bibfield  {journal} {\bibinfo  {journal}
  {Phys. Lett. B}\ }\textbf {\bibinfo {volume} {425}},\ \bibinfo {pages} {129}
  (\bibinfo {year} {1998}{\natexlab{a}})}\BibitemShut {NoStop}%
\bibitem [{\citenamefont {{H. Hiro-Oka and H.
  Minakata}}(1998{\natexlab{b}})}]{Hiro-oka_PLB_E}%
  \BibitemOpen
  \bibfield  {author} {\bibinfo {author} {\bibnamefont {{H. Hiro-Oka and H.
  Minakata}}},\ }\href@noop {} {\bibfield  {journal} {\bibinfo  {journal}
  {Phys. Lett. B}\ }\textbf {\bibinfo {volume} {434}},\ \bibinfo {pages}
  {461(E)} (\bibinfo {year} {1998}{\natexlab{b}})}\BibitemShut {NoStop}%
\bibitem [{\citenamefont {{H. Hiro-Oka and H. Minakata}}(2000)}]{Hiro-oka_PRC}%
  \BibitemOpen
  \bibfield  {author} {\bibinfo {author} {\bibnamefont {{H. Hiro-Oka and H.
  Minakata}}},\ }\href@noop {} {\bibfield  {journal} {\bibinfo  {journal}
  {Phys. Rev. C}\ }\textbf {\bibinfo {volume} {61}},\ \bibinfo {pages} {044903}
  (\bibinfo {year} {2000})}\BibitemShut {NoStop}%
\end{thebibliography}


%%%%%%%%%%%%%%%%%%%%%%%%%%%%%%%%%%%%%%%%%%%%%%%%%%%%%%%%%%%%%%%%%%%%%%%%%%%%%%%%%%%%%%%%%%%%%%%%%%%%%%%%%%%%%%%
\appendix
\section{Examples of squeeze operator}
\label{app:sqopexample}
In this appendix, we give three examples of the squeeze operator. %

The first example is given by the following $G(\vec{k},\vec{l})$: $G(\vec{k},\vec{l}) = g(\vec{k}) \ \delta (\vec{k} - \vec{l})$. 
The squeeze operator is obtained by substituting $G(\vec{k},\vec{l})$ directly:
\begin{equation} 
S = \exp \left( \frac{1}{2} \int d\vec{k} 
\left[ 
g^{*}(\vec{k}) \left( a(\vec{k}) \right)^2 
- 
g(\vec{k}) \left( a^{\dag}(\vec{k}) \right)^2 
\right]
\right).
\label{op:one_mode_like}
\end{equation}
This is just the operator given by Eq.~\eqref{eqn:unsatisfact_op}.
%%%%%%%%%%
The function $G(\vec{k},\vec{l})$ is represented as 
$G(\vec{k},\vec{l})= \left| g(\vec{k}) \right| e^{i\theta(\vec{k})} \delta(\vec{k}-\vec{l})$.
The (Bogoliubov) transformation is easily obtained by using Eq.~\eqref{eqn:sop}:
\begin{equation}
S^{\dag}(G) a(\vec{p}) S(G)  = 
\left( \cosh \left| g(\vec{p}) \right| \right) a(\vec{p})  
- e^{i\theta(\vec{p})}  \left( \sinh \left|  g(\vec{p}) \right| \right) a^{\dag}(\vec{p}) .
\end{equation}

%%%Case 2: 
The second example is given by the following $G(\vec{k},\vec{l})$: $G(\vec{k},\vec{l}) = g(\vec{k}) \ \delta (\vec{k} + \vec{l})$.
The function $G(\vec{k},\vec{l})$ is represented as 
$\left| g(\vec{k}) \right| e^{i\theta(\vec{k})}  \delta (\vec{k} + \vec{l})$.
The condition, $G(\vec{k},\vec{l}) = G(\vec{l},\vec{k})$, requires  $g(\vec{k}) = g(-\vec{k})$.
The squeeze operator is 
\begin{equation}
S = \exp \left( \frac{1}{2} \int d\vec{k} 
\left[ 
g^{*}(\vec{k}) a(\vec{k}) a(-\vec{k})
- 
g(\vec{k}) a^{\dag}(\vec{k}) a^{\dag}(-\vec{k})
\right]
\right).
\end{equation}
The coefficients 
$A_{2n}\left( \vec{p},\vec{k}_{2n} \right)$ and 
$B_{2n+1}\left( \vec{p},\vec{k}_{2n+1} \right)$ are 
\begin{subequations}
%%%%%%%
\begin{align}
&
A_{2n}\left( \vec{p},\vec{k}_{2n} \right) =
\left| g(\vec{p}) \right|^{2n} 
\delta(-\vec{p}+\vec{k}_{2n})
, \\ 
& 
B_{2n+1}\left( \vec{p},\vec{k}_{2n+1} \right) = 
\left| g(\vec{p}) \right|^{2n+1} 
e^{i \theta(\vec{p}) } \delta(\vec{p}+\vec{k}_{2n+1})
.
\end{align}
\end{subequations}
%%%%%%%%%%%%%%%%%%%
The transformation of the annihilation operator by the squeeze operator is 
\begin{equation}
S^{\dag}(G) a(\vec{p}) S(G)  = 
\left( \cosh \left| g(\vec{p}) \right| \right) a(\vec{p})  
- e^{i\theta(\vec{p})}  \left( \sinh \left|  g(\vec{p}) \right| \right) a^{\dag}(-\vec{p})  
. 
\end{equation}

%%%Case 3:
The final example is the squeeze operator described 
by smeared annihilation and creation operators.  
We use the smeared operators introduced by Andreev and Weiner \cite{contsqu}:
\begin{equation}
\tilde{a}_A(\vec{p}) = 
\int \frac{d\vec{k}}{(2\pi)^{3}} f_A(\vec{k}-\vec{p}) a_A(\vec{k}),
\label{eqn:a_tilde_in_mom}
\end{equation}
where $a_A(\vec{k})$ satisfies 
$[a_A(\vec{k}), a_A(\vec{l})] = (2\pi)^3 \delta(\vec{k}-\vec{l})$.
%%%%%%%%%%%%
The smearing function $f_A(\vec{x})$ is the Fourier transformation of $f_A(\vec{k})$. 
The integral of $|f_A(\vec{x})|^2$ is called the effective volume of particle source.

We attempt to construct the squeeze operator that 
is described by $\tilde{a}_A(\vec{p})$ and $\tilde{a}^{\dag}_A(\vec{p})$.
%%%%%%%%%% 
We take $G(\vec{k},\vec{l})$ as follows: 
\begin{equation}
G(\vec{k},\vec{l}) 
= \int \frac{d\vec{p}}{(2\pi)^3} \ h(\vec{p}) f_{A}^{*}(\vec{k}-\vec{p}) f_{A}^{*}(\vec{l}-\vec{p}).
\label{eqn:rel}
\end{equation}
The squeeze operator is obtained 
by substituting eq.~\eqref{eqn:rel} into the expression of the squeezed operator:
\begin{equation}
S = \exp \left( \frac{1}{2} \int d\vec{p}  \left[ 
h^{*}(\vec{p}) \left[ \tilde{a}_A(\vec{p})  \right]^{2}
- h(\vec{p}) \left[ \tilde{a}_A^{\dag}(\vec{p})  \right]^{2}
\right] \right)
.
\end{equation}
%%%%%%%%%%%%%%%%%%%%%%
Clearly,
the operator is obtained 
by replacing $a(\vec{k})$ and $a^{\dag}(\vec{k})$ in Eq.~\eqref{op:one_mode_like}
by $\tilde{a}_A(\vec{k})$ and $\tilde{a}^{\dag}_A(\vec{k})$ respectively.

%%%%%%%%%%%%%%%%%%%%%%%%%%%%%%%%%%%%%%%%%%%%
\section{Equivalence of two squeezed states} 
\label{app:sqop}
A squeezed state is defined as $|{\rm sq}\rangle = D(\alpha) S(G) |0\rangle$. 
Contrarily, another definition exists, $|\overline{\rm sq}\rangle = S(G)D(\beta)|0\rangle$. 
In this section, we attempt to show the relation between these two states 
when $G(\vec{k},\vec{l}) = g(\vec{k}) g(\vec{l})$.

We prove here that the state $|\overline{\rm sq}\rangle$ is an eigen state of 
the operator $b(\vec{p})$ that is defined by $b(\vec{p})=S(G) a(\vec{p}) S^{\dag}(G)$. 
By operating $b(\vec{p})$ to the state $|\overline{\rm sq}\rangle$, we obtain
\begin{align}
b(\vec{p}) |\overline{\rm sq}\rangle =
b(\vec{p}) S(G) D(\beta) |0\rangle = S(G) a(\vec{p}) D(\beta)|0\rangle .
\end{align}
%%%
It is shown that $|\overline{\rm sq}\rangle$ is an eigen state of the operator $b(\vec{p})$
by using the relation,
$D^{\dag}(\beta) a(\vec{p})  D(\beta)  = a(\vec{p}) + \beta(\vec{p})$:
\begin{equation}
b(\vec{p}) |\overline{\rm sq}\rangle  = \beta(\vec{p}) |\overline{\rm sq}\rangle .
\label{eqn:bsqtilde}
\end{equation}

Next, we consider the state $b(\vec{p}) |{\rm sq}\rangle$: 
%%%%%%%%%%%%%%%%%%%%
\begin{align}
b(\vec{p}) |{\rm sq}\rangle =& 
\left[
\alpha(\vec{p}) 
+ \left( \frac{\cosh L -1}{L} \right) g(\vec{p}) \left( \int d^{3}\vec{k} \ g^{*}(\vec{k}) \alpha(\vec{k}) \right)
+ \left( \frac{\sinh L}{L} \right) g(\vec{p}) \left( \int d^{3}\vec{k} \ g(\vec{k}) \alpha^{*} (\vec{k}) \right)
\right] |{\rm sq}\rangle .
\label{eqn:btimessq}
\end{align}
%%%%%%%%%%%%%%%%%%%%
Therefore 
$|{\rm sq}\rangle$ is identical to $|\overline{\rm sq}\rangle$,
when the following condition is satisfied: 
\begin{equation}
\beta(\vec{p}) = 
\alpha(\vec{p}) 
+  \left( \frac{\cosh L - 1}{L} \right) g(\vec{p}) \left( \int d^{3}\vec{k} \ g^{*}(\vec{k}) \alpha(\vec{k}) \right)
+  \left( \frac{\sinh L}{L} \right) g(\vec{p}) \left( \int d^{3}\vec{k} \ g(\vec{k}) \alpha^{*}(\vec{k}) \right) .
\end{equation}

\end{document}